\documentclass[12pt,preprintnumbers, nofootinbib]{revtex4}

\usepackage{amssymb}                   %Needed for certain symbols
\usepackage{amsmath}                    % Needed for multline
\usepackage{fancyhdr}                     % Needed for title page headings
\usepackage{graphicx}                     % Needed for includegraphics
\usepackage{hyperref}                     % Needed for hyper references
\usepackage{verbatim}                     % Needed for using latex as a typewriter, and for comment!
\usepackage{subfigure}                    % Needed for using subfigures

\numberwithin{equation}{section}      %Needed for equation numbers to follow sections

\newcommand{\D}{\Delta}
\newcommand{\G}{\Gamma}
\renewcommand{\L}{\Lambda}
\renewcommand{\l}{\lambda}
\newcommand{\g}{\gamma}
\renewcommand{\O}{\mathcal{O}}
\newcommand{\s}{s_\theta}
\renewcommand{\c}{c_\theta}
\renewcommand{\bf}[1]{\textbf{#1}}
\newcommand{\bra}{\langle}
\newcommand{\ket}{\rangle}

\newcommand{\sm}{\sqrt{\hat s_{max}}}
\renewcommand{\b}[1]{\mathbf{#1}}

\begin{document}

\title{Bounds from LEP on unparticle interactions with electroweak bosons.}

\author{Scott Kathrein}
\email{kathrein@physics.rutgers.edu}
\affiliation{Department of Physics and Astronomy,
Rutgers University, Piscataway, NJ 08854}

\author{Simon Knapen}
\email{knapen@physics.rutgers.edu}
\affiliation{Department of Physics and Astronomy,
Rutgers University, Piscataway, NJ 08854}

\author{Matthew J. Strassler}
\email{strassler@physics.rutgers.edu}
\affiliation{Department of Physics and Astronomy,
Rutgers University, Piscataway, NJ 08854}

\preprint{RUNHETC-2010-28}

\begin{abstract}
A conformally invariant hidden sector is considered, with a scalar
operator $\O$ of low dimension that couples to the electro-weak gauge
bosons of the Standard Model, via terms such as
$F^{\mu\nu}F_{\mu\nu}\O$. By examining single photon production at
LEP, we bound the strength of these interactions. We apply our
results, along with those of Delgado and Strassler \cite{ds09} and of
Caracciolo and Rychkov \cite{cr09}, to improve the bound on $4\gamma$
production through ``unparticle self-interactions'', as proposed by Feng et al.
\cite{frt08}.  We find the maximum allowable cross-section is of order
a few tens of femtobarns at the 14 TeV LHC, and lies well below 1 fb
for a wide range of parameters.
\end{abstract}

\maketitle

\section{Introduction}

A ``hidden'' sector of light particles, none of
which carry standard model quantum numbers, is still allowed by
experiment.
Neither direct searches, nor indirect tests of the
standard model, nor cosmology or astrophysics can exclude this
possibility.  If the coupling of such a sector to our own
is purely through gravitation, constraints are extremely weak.  
But if additional interactions, with
characteristic energy scales far below the Planck scale, are present, then
it is possible to obtain some correlated constraints on the strength of those interactions and
the contents of the hidden sector.

Since the contents of such a sector are all neutral and may all be
stable or metastable, production of anything in that sector may generally
be invisible. In such a case, constraints may be obtained at a wide
range of particle colliders, using their searches for unexplained
sources of missing momentum.  At a hadron collider, the typical search
is for a jet or a photon plus missing transverse momentum.  At an
electron-positron collider, a powerful constraint may be obtained from
searches for ``photon-plus-nothing'' --- events in which a photon
is observed
whose momentum is not balanced against any visible object.  Since the
collision energy and momentum are known at a lepton collider, the
four-momentum of the missing object, and its square, the ``missing
mass,'' may be reconstructed.  The events are very clean and easy to
interpret, though a background from $e^+e^-\to \gamma\nu\bar\nu$, where
the neutrinos may or may not originate from an on-shell $Z^0$, must be
removed.

In this article, we consider constraints on hidden sectors from
photon-plus-nothing searches at the Large Electron-Positron (LEP) collider
in its two stages, LEP I (at the $Z$ boson peak) and LEP II (at
center-of-mass energies up to 209 GeV).  Our focus here will be on
exactly or approximately conformal hidden sectors, now often called
``unparticle'' sectors \cite{Georgi:2010zz}. We will obtain
constraints on couplings of $SU(2)\times U(1)$ gauge bosons to
low-dimension scalar operators in such sectors.  We only consider operators with dimension less than 2. 
(For $\D > 2$, operator renormalizations become necessary and the calculations 
become sensitive to the ultraviolet, leaving them less predictive. 
Note also that unitarity requires $\D\geq1$.)

As an application of
our results, we will combine them with the work of \cite{ds09} and \cite{cr09}
to obtain limits on the process $gg\to \gamma\gamma\gamma\gamma$,
highlighted in \cite{frt08} as a possible source of a large effect of an
unparticle sector.  We will see that where qualitatively
new constraints can be obtained,
the allowed signals must lie below 5 fb, even at a 14 TeV collider.

In section 2, we will discuss the general theoretical background
and calculations needed for this paper.  In section 3, we will obtain
bounds from LEP results.  Finally, we will apply these bounds in the
particular case of four-photon production at the LHC.
 
\section{Nature of the CFT Coupling\label{sectiontheory}}
\label{sec:natureCFT}

In what follows, we imagine that, through new physics
somewhat above the TeV scale, a hidden conformal (unparticle) sector
is coupled to the standard model gauge bosons.  (Couplings to fermions
risk flavor-changing neutral currents, unless they occur through
conserved currents of dimension 3, in which case contact terms
generally dominate \cite{gir08}.)  We assume the following
Lagrangian, where a scalar primary operator $\O$ of the conformal
sector couples to the electroweak gauge fields.
\begin{equation}
\label{firstlambdadef}
\delta\mathcal{L} = \frac{\l_1}{\L_1^\D}  B^{\mu\nu}B_{\mu\nu} \O
+ \frac{\l_2}{\L_2^\D}  W_a^{\mu\nu}W^a_{\mu\nu}\O.
\end{equation}
Here, $|\l_1|=|\l_2|=1$ and $\L_1$ and $\L_2$ are real and positive. The
two conformal operators in this expression are assumed to be the same,
with scaling dimension $\D$; we consider only $2\geq \D\geq 1$.  
As is the standard operating procedure
in the literature on scalar unparticle sectors, we ignore serious
subtleties involving the generation of the operator $|\O|^2$ through
quantum effects, assuming that (as for the Higgs mass operator) the
coefficient of the operator is suppressed through an unspecified
mechanism.  Examples of possible mechanisms include supersymmetry;
see for example \cite{nelson}.

After electroweak symmetry breaking
mixes the $B$ and the $W^3$ to form the photon and the $Z^0$,
the Lagrangian contains the terms 

\begin{align}
\delta\mathcal{L}_{\g\g} &= 2(\c^2\frac{\l_1}{\L^\D_1}+\s^2\frac{\l_2}{\L^\D_2}) (\partial_\mu A_\nu \partial^\mu A^\nu - \partial_\mu A_\nu \partial^\nu A^\mu) \O\\
\delta\mathcal{L}_{Z\g} &= 4\c \s(\frac{\l_2}{\L^\D_2}-\frac{\l_1}{\L^\D_1}) (\partial_\mu A_\nu \partial^\mu Z^\nu - \partial_\mu A_\nu \partial^\nu Z^\mu) \O\\
\delta\mathcal{L}_{ZZ} &= 2(\s^2\frac{\l_1}{\L^\D_1}+\c^2\frac{\l_2}{\L^\D_2}) (\partial_\mu Z_\nu \partial^\mu Z^\nu - \partial_\mu Z_\nu \partial^\nu Z^\mu) \O\\
\delta\mathcal{L}_{WW} &= 2\frac{\l_2}{\L_2^\D} (\partial_\mu W^\pm_\nu \partial^\mu W^{\pm\nu} - \partial_\mu W^\pm_\nu \partial^\nu W^{\pm^\mu})\O,
\end{align}
where $\c\equiv \cos(\theta_w)$ and $\s\equiv \sin(\theta_w)$. The following definitions will simplify formulae
\begin{align}
\frac{\l_\g}{\L_\g^\D} &\equiv \c^2\frac{\l_1}{\L^\D_1}+\s^2\frac{\l_2}{\L^\D_2} \\
\frac{\l_Z}{\L_Z^\D} &\equiv \frac{\l_2}{\L^\D_2}-\frac{\l_1}{\L^\D_1}
\end{align}
with $|\l_\g|=|\l_Z|=1$ and $\L_\g$ and $\L_Z$ real and positive.
For the present article the most interesting interactions will be those involving the photon. 
The photon--photon--unparticle vertex, and the photon--$Z^0$--unparticle vertex, lead to
vertices with Feynman rules \cite{cky07,frt08}
\begin{align}
\g\g\O &\rightarrow -4i\frac{\l_\g}{\L_\g^\D}(g^{\mu_1\mu_2}k_1\cdot k_2 - k_1^{\mu_2} k_2^{\mu_1}) \\
Z\g\O &\rightarrow -4i\c \s\frac{\l_Z}{\L_Z^\D}(g^{\mu_1\mu_2}k_1\cdot k_2 - k_1^{\mu_2} k_2^{\mu_1}).
\end{align}

In Sec.~\ref{sec:fourphotons},
we will assume that the gluons couple to the unparticle sector as well.
We will rename $\O$ and $\D$ as $\O_\g$ and $\D_\g$, and permit
the gluons to couple to an operator $\O_g$
\begin{align}
\delta\mathcal{L} &= \frac{\l_g}{\L_g^\D} G^a_{\mu\nu}G_a^{\mu\nu} \O_g\\
&= 2\frac{\l_g}{\L_g^\D} (\partial_\mu G^a_\nu \partial^\mu G_a^\nu - \partial_\mu G^a_\nu \partial^\nu G_a^\mu)\O_g \ .
\end{align}
Here, $\O_g$ (and $\D_g$) may or may not be the same as $\O_\g$ (and
$\D_\g$).  (Note that \cite{frt08}, in considering four
photon production at the Tevatron and LHC,
assumed $\O_g=\O_\g$.)
This Lagrangian yields the vertex
\begin{equation}
gg\O_g\rightarrow -4i\frac{\l_g}{\L_g^{\D_g}} (g^{\mu_1\mu_2} k_1 \cdot k_2 - k_1^{\mu_2} k_2^{\mu_1}) \delta_{a_1}^{a_2}.
\end{equation}

\section{Cross Section}

\begin{figure}[t]\centering
\subfigure[Photon mediated\label{photonmediated}]{\includegraphics[width=0.3\textwidth]{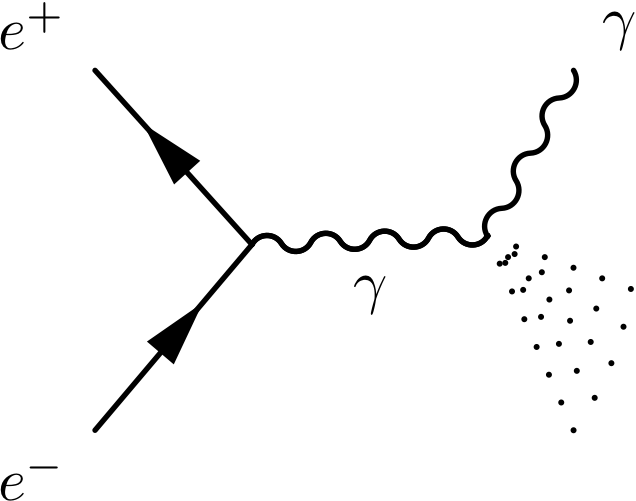}}\hspace{3cm}
\subfigure[Z mediated\label{Zmediated}]{\includegraphics[width=0.3\textwidth]{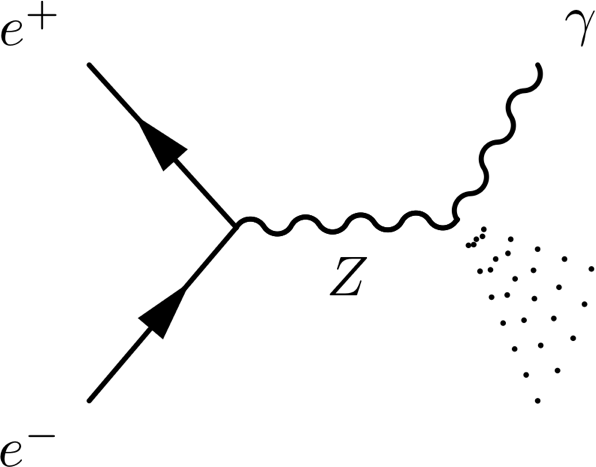}}
\caption{Feynman diagrams for production of hidden states in the LEP collider. The dots represent states in the conformal hidden sector.\label{feyndiag}}
\end{figure}

The amplitude for $e^+e^- \rightarrow \{\g \textrm{ or } Z^0\} \rightarrow\g \O$ at tree level (figure \ref{feyndiag}) is 
\begin{align}
\overline{\sum} |M|^2 &= A(\L) \frac{e^2}{s}(t^2+u^2) \\
\end{align}
where
\begin{align}
\label{alambda}
A(\L) &\equiv \left(A_Z\frac1{\L_Z^{2\D}} + A_\g\frac1{\L_\g^{2\D}} + A_{Z\g}\frac1{\L_Z^\D \L_\g^\D}\right)\\
A_Z &\equiv (\frac12-2\s^2+4\s^4) \left(\frac{s^2}{(s-m_Z^2)^2+m_Z^2\G_Z^2}\right)\\
A_\g &\equiv 4\\
\label{deltadef}
A_{Z\g} &\equiv 2(1-4\s^2) \left(\frac{(s-m_Z^2)\cos(\delta) - m_Z\G_Z\sin(\delta)}{(s-m_Z^2)^2+m_Z^2\G_Z^2}\right) s.
\end{align}
Here $\delta$ is the relative phase difference between the two
diagrams in \ref{photonmediated} and \ref{Zmediated}, originally 
parametrized by $\l_\g$ and $\l_Z$. 
The result for $A_\g$ matches \cite{fds03} appropriately in the $\D\rightarrow 1$ limit.

%
% begin figure
%end figure
%

The differential cross section is calculated with respect to the Mandelstam variables $t$ and $u$, as well as with respect to $ \cos\theta$ and $q$, with $q$ the energy of the final state photon. 
\begin{align}
\frac{d^2\sigma}{dt\:du} &= \frac{(4\pi)^{1-2\D}}{4\G(\D-1)\G(\D)} A(\L) e^2 \frac{(t^2+u^2)(s+t+u)^{\D-2}}{s^3} \\
\frac{d^2\sigma}{dq\:d\!\cos\theta} &= \frac{(4\pi)^{1-2\D}}{\G(\D-1)\G(\D)} A(\L) e^2 q^3 s^{\D-3} (1-2\frac{q}{\sqrt{s}})^{\D-2} (1+\cos^2\theta)\label{photoncrosssection}
\end{align}
The latter result is most suitable for numerical integration to compare to experiments with lepton colliders. 

 At the peak of the $Z^0$ resonance, $|A_Z| \sim
85 |A_\g+A_{Z\g}|$, for $\delta=0$ or $\delta=\pi$. This ratio becomes
the smallest for $\delta=3\pi/2$, where $|A_Z| \sim 35
|A_\g+A_{Z\g}|$. From LEP I data at the $Z^0$ resonance, we mainly
obtain a bound of $\L_Z$ alone. At LEP II energies, near 200 GeV,
$|A_\g| \sim 7 |A_Z+A_{Z\g}|$ for $\delta=0$, up to a maximum of
$|A_\g| \sim 19 |A_Z+A_{Z\g}|$ for $\delta=\pi$, and thus we obtain a
limit mainly on $\L_\g$. 

\section{Bounds from LEP data}

\subsection{From LEP I data}

During the first run of the LEP experiment, data was collected at the
Z-resonance. Unparticle production is therefore dominated by the $A_Z$
term (see Eq.~\ref{alambda}), as was argued in the previous
paragraph.  To obtain a worst case bound on $\L_Z$, we will neglect
contributions from the photon channel.\footnote{Strictly speaking, if
the phase $\delta$ is such that interference is maximally destructive,
including the photon channel can decrease the signal by up to $1\%$.
But this is less than other systematic errors discussed in
\ref{SectionError}, so we neglect it.}  This bound could be only
slightly improved by incorporating the data from LEP II.

As can be seen from the energy distribution of the single photon in
formula \ref{photoncrosssection}, unparticles tend to produce very
hard photons for values of $\D$ less than two. The Standard Model
background for this signal on the other hand is only of order 0.5-1
events. To obtain optimal sensitivity for our bounds we require the
photon energy to be larger than a certain minimum energy, $E_{cut}$,
which is determined by optimizing the sensitivity for the bound on
$\L_Z$. More details on the energy cuts can be found in appendix
\ref{analysisappendix}. None of the four LEP I detectors observed
events that pass our energy cuts
\cite{lep1opal,lep1l3,lep1delphi,lep1aleph}.

Combining (as described
in the appendix) the available
data from all four experiments we establish a 95\% confidence level (CL) 
bound on $\L_Z$,
following \cite{poisson}. Our bounds are displayed in table
\ref{table1}. A plot of 
the allowed regions for $\L_2$ and $\L_1$, the couplings to the $SU(2)\times
U(1)$ bosons, is also given
in figure \ref{fig1}.   For this plot, the entire matrix element was
taken into account. 

The value we give for $\L_Z$ when $\D\to 1$ is
consistent with the known branching fraction for $Z\to \gamma+X$,
where $X$ is a very light new invisible particle and $E_\gamma\sim 45$
GeV.  The partial width for this process would be
\begin{equation}
\G=\frac{\c^2\s^2 M_Z^3}{6\pi\L_Z^2}.
\end{equation}
Since no 45 GeV photons plus missing energy were observed in any of
the four LEP experiments, one can obtain a model-independent $95\%$ CL
bound on the branching ratio.  The best such published bound,
$1.1 \times 10^{-6}$, was obtained by the L3 experiment \cite{lep1l3}, and
this can be converted to $\L_Z > 51$ TeV with $95\%$ CL. The bound in
our table above is consistent with this, though somewhat stronger
since we combine all four LEP I experiments in our calculation.

\subsection{From LEP II data}

The second run of LEP scanned center of mass energies from 130 GeV to
209 GeV. Since the cross section \ref{photoncrosssection} grows with
$\sqrt{s}$, the highest collider energies will give us the best
bounds. The dominant mode of unparticle production at these energies
is via the photon channel, and interference effects are small, so we
obtain a worst case bound on $\L_\g$ by neglecting contribution from
the $Z$ channel.

As mentioned above, the best bounds on $\L_\g$ can be obtained from the highest energies
at LEP II.  Our bounds below therefore take account only of
data from energy in the range 183-209 GeV.  In particular,
DELPHI \cite{lep2delphi}, ALEPH \cite{lep2aleph}, and L3
\cite{lep2l3} published results for $\sqrt{s}$ between 183 GeV and 209
GeV, while OPAL \cite{lep2opal} did not publish a result above 189 GeV. If one 
accounted for the results at lower collider energies, it
would be possible to extract a bound that is slightly better than
ours.

Since the collider energy was changed over time, the data in
\cite{lep2opal,lep2delphi,lep2l3,lep2aleph} are given not in terms of
the photon energy itself but in terms of the ``missing mass'', the
mass that an invisible particle would have had if it were recoiling
from the observed photon.  For low $\D$, the signal is peaked in the
low missing mass region,\footnote{$\D$=1 corresponds to a massless
invisible scalar particle.}  while for higher $\D$, the signal is
rather flat. Since the standard model background is smallest in the
low missing-mass region, far from the $Z\to\nu\bar\nu$ peak,
integrating the signal from zero missing mass up to some maximum
missing-mass $M_{cut}$ yields the best bounds.  The selection of
$M_{cut}$ for each $\D$, and other details of our analysis, are
described in appendix \ref{analysisappendix}.  OPAL, ALEPH and in
particular DELPHI detected several events that pass our energy cuts.
The bounds we obtain are found in table \ref{table1}. 
The allowed regions for
$\L_2$ and $\L_1$ (the couplings to the $SU(2)\times U(1)$ bosons)
are given in figure \ref{fig2}; here the entire matrix
element including the $Z$ contribution is taken into account.

Finally, we wish to note that there are ambiguities regarding the interpretation
of certain published plots which affect the analyses, and require us to make
certain assumptions. A key ambiguity regarding our analysis revolves around the
result of the DELPHI experiment.  (At DELPHI, as with the other
experiments, we only use data from approximately 45 to 135 degrees; 
see appendix \ref{analysisappendix}.) In the bin at zero missing mass,
there are 7 events, above one expected in background.  This bin
was used as an underflow, and at least 6 of the events\footnote{We are
  very grateful to C.~Mateuzzi of DELPHI for providing us with
  considerable information about these events.} are from ``photons''
with energy larger than half the beam energy, giving a negative
apparent missing mass, which is not consistent with our signal. 
We therefore view the interpretation of one unexplained
event in this bin as ambiguous.  There are several choices, including discarding this
bin as having large background, discarding the 7th event in the bin as being more
plausibly background than signal, discarding the DELPHI data completely, etc.

Our table above reflects the most liberal (but in our view, also the most
plausible) assumption that the seventh event in the zero-missing-mass bin is,
like the other six, from a background source.  It is likely that this
could be shown to be the case with sufficient information
about the DELPHI data.  If instead we treat the seventh event as a potential
signal, the effect on our bounds is substantial in the regime where $\D$ is small,
on the order of 20\% in $\L_\g$.

\begin{table}
\begin{center}
\begin{tabular}{|l||r@{.}l r@{.}l|}
 \hline
 \textbf{\em $\D$} 
    &\multicolumn{2}{c}{ \textbf{$\L_Z$} }
    &\multicolumn{2}{c|} {\textbf{$\L_\g$} }
    \\ \hline\hline
 1& 69&5    &  25&2\\
 1.01& 59&0 &  23&0\\
 1.05& 40&7 &  13&2\\
 1.1&   26&6&  8&0\\
 1.2&  12&7&  3&6\\
 1.3&   6&8&  2&0\\
 1.4&  4&0&  1&2\\
 1.5&   2&5&  0&79         \\
 1.6&    1&6&  0&57\\
 1.7&   1&1&  0&41\\ 
 1.8&    0&80&  0&30\\
 1.9&      0&60&  0&24\\ 
 2&     0&46& 0&19\\ \hline
\end{tabular}
\caption{95 $\%$ confidence level lower bounds on the given scales, 
in TeV, from LEP data. 
For bounds on $\Lambda_1$ and $\Lambda_2$, see the figures in
Appendix A.\label{table1}}
\end{center}
\end{table}

\subsection{Error Estimate\label{SectionError}}

The largest uncertainty in both the LEP I and the LEP II
analyses (other than the ambiguities in the LEP II data described
above) is due to the systematic errors in manually reading the
backgrounds from the graphs. However, in the case of LEP I, this error
only contributes in the calculation of the cuts, as no events are
found in the signal region \cite{poisson}. Furthermore we find that the bounds are
not very sensitive to cuts, and the error due to the background only
contributes a few percent to the total error on the bounds. When
accounting for experimental uncertainties we can estimate the total
uncertainty on the bounds to be within 5\%.

For LEP II, the systematic uncertainty from reading
backgrounds from the plots is significantly larger. Moreover the bounds do
depend directly on the background in this case, although the
dependence is very mild. The total uncertainties on the bound on
$\L_\g$ are estimated to be smaller than 10\%.  In these estimates we
ignore the much larger systematic uncertainties that arise from the 
ambiguities described above in the interpretation of the
published data.

\section{Four Photon Signals}
\label{sec:fourphotons}
Multi-point correlation functions (sometimes called ``unparticle
self-interactions'') for the conformal operators $\O$ have been
proposed as a possible source at the LHC of very large new-physics
signals --- including four-photon signals as large as 10 nb
\cite{frt08}.  But as shown in \cite{ds09}, CDF limits on signals that
give a jet plus missing transverse momement (MET), and general
considerations of unitarity and self-consistency, strongly constrain
such processes, to a few fb in some regimes (including those
considered in \cite{frt08}) and a few pb in some other regimes.  The
results of the current paper, combined with work of \cite{cr09}, allow
us to improve constraints by several orders of magnitude.  Limits on
the maximum cross-section for $gg\to 4\gamma$ at a 14 TeV LHC are
given in table \ref{4photonbounds}.  In this table, we assume
that the standard model gauge bosons couple to operators $\O_g$ and
$O_\g$, with dimensions $\D_g$ and $\D_\g$, as described in Sec.~\ref{sec:natureCFT}.

Before explaining how we obtained these results, let us make a couple
of brief comments.  Compared to \cite{ds09}, our new bounds for
$\Delta_\gamma<1.7$ are far stronger, especially for small
$\Delta_\gamma$, by as many as five orders of magnitude. We can see
that bounds are below 5 fb for $\D_\g<1.7$.  For $\D_\g>1.7$ we must
rely on the methods of \cite{ds09} (extended to 14 TeV), obtaining
constraints of a few tens of fbs or less at low to moderate $\D_g$. We
should note also that the bounds at low $\D_g$ are obtained from a CDF
jet-plus-MET measurement \cite{cdf} that uses only 1.1 inverse fb of
data, much less than the total Tevatron data set. 

\begin{table}[b]
\resizebox{\textwidth}{!}{\begin{tabular}{|c||r@{.}l|r@{.}l|r@{.}l|r@{.}l|r@{.}l|r@{.}l|r@{.}l|r@{.}l|r@{.}l|r@{.}l|r@{.}l|}\hline
 $\phantom{a}_{\D_g} \diagdown^{\D_\g}$ & \multicolumn{2}{c|}{1.05} &\multicolumn{2}{c|}{ 1.1} &\multicolumn{2}{c|}{ 1.2} &\multicolumn{2}{c|}{ 1.3} &\multicolumn{2}{c|}{ 1.4} &\multicolumn{2}{c|}{ 1.5} & \multicolumn{2}{c|}{1.6} &\multicolumn{2}{c|}{ 1.7} &\multicolumn{2}{c|}{ 1.8} &\multicolumn{2}{c|}{ 1.9} &\multicolumn{2}{c|}{ 2.0}\\\hline\hline
 1.05 & 2&$7\times 10^{\text{-6}}$   & $2$&$7\times 10^{\text{-5}}$  & $4$&$8\times 10^{\text{-4}}$ &0&010& 0&093& 0&62 & \bf{1}&\bf{1  } & \bf{1}&\bf{7  } & \emph{3}&\emph{8 }& \emph{2}&\emph{3 }& \emph{1}&\emph{4} \\\hline
 1.1   & 5&$1\times 10^{\text{-6}}$   & $5$&$2\times 10^{\text{-5}}$  & $6$&$7\times 10^{\text{-4}}$ &0&014 & 0&13& 0&89 & \bf{1}&\bf{4   }  &\bf{1}&\bf{6}& \emph{9}&\emph{6 }& \emph{5}&\emph{9 }& \emph{3}&\emph{7} \\\hline
 1.2   & 1&$5\times 10^{\text{-5}}$   & $1$&$4\times 10^{\text{-4}}$  & $1$&$3\times 10^{\text{-3}}$&0&023 &0&37& 2&4   & \bf{2}&\bf{3}   & \bf{1}&\bf{7}   & \emph{2}&\emph{3 }& \emph{1}&\emph{4 }& \emph{7}&\emph{1} \\\hline
 1.3   & 3&$7\times 10^{\text{-5}}$     & $\b{2}$&$\b{8\times 10^{\b{-4}}}$  & $\b{3}$&$\b{2\times 10^{\b{-3}}}$ &\bf{ 0}&\bf{031} &\bf{ 0}&\bf{33}   & \bf{1}&\bf{7}  & \bf{1}&\bf{2}   & \bf{0}&\bf{91} & \emph{16}&& \emph{9}&\emph{3 }& \emph{5}&\emph{4} \\\hline
 1.4   & $\b{3}$&$\b{3\times 10^{\b{-5}}}$   & $\b{2}$&$\b{5\times 10^{\b{-4}}}$  & $\b{2}$&$\b{3\times 10^{\b{-3}}}$ &\bf{ 0}&\bf{023} &\bf{ 0}&\bf{24}   & \bf{1}&\bf{2}   & \bf{0}&\bf{73} & \bf{0}&\bf{56} & \emph{12}&& \emph{7}&\emph{1 }&\emph{ 4}&\emph{5} \\\hline
 1.5   & $\b{3}$&$\b{6 \times 10^{\b{-5}}}$  & $\b{2}$&$\b{4\times 10^{\b{-4}}}$  & $\b{2}$&$\b{8\times 10^{\b{-3}}}$ &\bf{ 0}&\bf{025 }& \bf{0}&\bf{19}   & \bf{0}&\bf{78} & \bf{0}&\bf{57 }& \bf{0}&\bf{37} & \emph{9}&\emph{3 }& \emph{5}&\emph{4 }& \emph{3}&\emph{2} \\\hline
 1.6   & $\b{3}$&$\b{6\times 10^{\b{-5}}}$   & $\b{2}$&$\b{6\times 10^{\b{-4}}}$  & $\b{2}$&$\b{3\times 10^{\b{-3}}}$ & \bf{0}&\bf{021} &\bf{ 0}&\bf{16}   & \bf{0}&\bf{55 }& \bf{0}&\bf{48} & \bf{0}&\bf{31} & \emph{7}&\emph{1 }& \emph{4}&\emph{7 }& \emph{2}&\emph{5 }\\\hline
 1.7   & $\b{4}$&$\b{7 \times 10^{\b{-5}}}$  & $\b{2}$&$\b{9\times 10^{\b{-4}}}$  & $\b{2}$&$\b{7\times 10^{\b{-3}}}$ & \bf{0}&\bf{024} &\bf{ 0}&\bf{16}   &\bf{ 0}&\bf{50} & \bf{0}&\bf{35 }& \bf{0}&\bf{26} & \emph{5}&\emph{4 }& \emph{3}&\emph{2 }& \emph{2}&\emph{0} \\\hline
 1.8   & $\b{4}$&$\b{4 \times 10^{\b{-5}}}$  & $\b{2}$&$\b{2\times 10^{\b{-4}}}$  & $\b{1}$&$\b{7\times 10^{\b{-3}}}$ & \bf{0}&\bf{022} & \bf{0}&\bf{20}   & \bf{0}&\bf{38} & \bf{0}&\bf{32} & \bf{0}&\bf{23}& \emph{4}&\emph{2 }& \emph{2}&\emph{5}& \emph{1}&\emph{5} \\\hline
 1.9   & $\b{3}$&$\b{4 \times 10^{\b{-5}}}$  & $\b{1}$&$\b{6\times 10^{\b{-4}}}$  & $\b{1}$&$\b{5\times 10^{\b{-3}}}$ &\bf{ 0}&\bf{014} &\bf{ 0}&\bf{15}   & \bf{0}&\bf{36} & \bf{0}&\bf{29} & \bf{0}&\bf{23} & \emph{3}&\emph{2 }& \emph{2}&\emph{0 }&\emph{ 1}&\emph{2} \\\hline
 2.0   & $\b{2}$&$\b{7 \times 10^{\b{-5}}}$  & $\b{1}$&$\b{3\times 10^{\b{-4}}}$  & $\b{8}$&$\b{7\times 10^{\b{-4}}}$ &\bf{ 0}&\bf{013} &\bf{ 0}&\bf{14}   &\bf{0}&\bf{35} &\bf{ 0}&\bf{31}   & \bf{0}&\bf{24  } & \emph{2}&\emph{5   }& \emph{1}&\emph{5 }& \emph{0}&\emph{96}\\\hline
\end{tabular}}
\caption{Bounds on 4 photon production, in fb. Values in regular font are obtained using only
experimental limits on $\Lambda_g$ and $\Lambda_\g$; see also Appendix B.  
Values in boldface are obtained from experimental and unitarity bounds, or unitarity bounds only, on these
scales. The values in italics are calculated using the unitarity argument of \cite{ds09}. \label{4photonbounds}}
\end{table}
\subsection{Obtaining the bounds}

In general, the cross section for $gg\to 4\gamma$ is proportional to
${C_3^2}{\L_g^{-2\D_g}\L_\g^{-4\D_\g}}\hat s^{\D_g+2\D_\g-1}$, where
$C_3$ is the coefficient of the three-point function
$\bra\O_g\O_\g\O_\g\ket$, and the scales $\L_g,\L_\g$ and dimensions
$\D_g,\D_\g$ are as defined in section \ref{sectiontheory}.  (In \cite{frt08} both the
gluons and the photons are assumed to couple to the same operator in
the conformal sector, but this is an unnecessary assumption.)  The
potentially enormous cross-sections suggested by \cite{frt08} arise
from the rapid growth with $\hat s$; even strong limits on $4\gamma$
production at the Tevatron naively allow very large LHC signals.  But
\cite{frt08} did not consider unitarity, or direct and indirect
constraints on $\Lambda_g$, $\Lambda_\gamma$ and $C_3$.  In \cite{ds09},
experimental
and theoretical bounds on $\Lambda_g$ were found (table \ref{gluonbound}), along
with a simple unitarity argument that eliminated the possibility of
very large cross-sections.  In the current article we have found
experimental bounds on $\Lambda_\gamma$, which (as described below) we
may supplement with theoretical bounds.  And recently, unitarity
constraints on $C_3$, from internal consistency arguments of the
conformal field theory, were obtained in \cite{cr09} for
$\D_\gamma<1.7$ and any $\D_g$.  We now explain how these bounds are
obtained and combined together into table \ref{4photonbounds}.

\begin{table}
\begin{center}
\begin{minipage}{0.3\textwidth}
\begin{tabular}{|c|c|}
 \hline
 $\D_g$ \ & \ $\L_g \ ({\rm TeV})$\\ \hline
1.05 \ & \  9.19
\\ 1.10 \ & \  6.82
\\ 1.15 \ & \  5.18
\\ 1.20 \ & \  4.03
\\ 1.25 \ & \  3.19
\\ 1.30 \ & \  2.58
\\ 1.35 \ & \  2.11
\\ 1.40 \ & \  1.75
\\ 1.45 \ & \  1.48
\\ 1.50 \ & \  1.26\\\hline
\end{tabular}
\end{minipage}\hspace{0.55cm}
\begin{minipage}{0.3\textwidth}
\begin{tabular}{|c|c|}\hline
$\D_g$ \ & \ $\L_g \ ({\rm TeV})$\\ \hline
 1.55 \ & \  1.08
\\ 1.60 \ & \  0.94
\\ 1.65 \ & \  0.82
\\ 1.70 \ & \  0.73
\\ 1.75 \ & \  0.64
\\ 1.80 \ & \  0.58
\\ 1.85 \ & \  0.52
\\ 1.90 \ & \  0.47
\\ 1.95 \ & \  0.43 \\ \hline
\multicolumn{2}{c}{} \\
\end{tabular}
\end{minipage}
\caption{Lower bounds (quoting and extending the results of
\cite{ds09}) on the interaction scale $\Lambda_g$ as a function of
$\D_g$, using only constraints from jet-plus-MET studies at CDF
\cite{cdf}.  The unitarity considerations also discussed in
\cite{ds09} are not applied here.
\label{gluonbound}}
\end{center}
\end{table}

In the regime $\D_\g>1.7$, indicated by numbers in italics in the
table, the constraints obtained in \cite{ds09} are extended to a 14
TeV LHC, using bounds on $\L_g$ only.
Direct experimental bounds on $\Lambda_g$ arise because the
gluon-gluon-unparticle interaction can generate a large
jet-plus-MET  signature
\cite{ds09}.  Limits from CDF \cite{cdf} using 1.1 $\mathrm{fb}^{-1}$ of data
(unfortunately not yet updated for the current, much larger, Tevatron
data sets) were obtained in \cite{ds09}, and are extended in
table \ref{gluonbound}. These bounds are powerful at
small $\D_g$. 

A theoretical bound on $\L_g$ is obtained as follows.  
A coupling of gluons of the form $G^2\O$ corrects the
$\bra\O(p)\O(-p)\ket$ two-point function by a computable amount.  Once
this correction becomes large enough that the two-point function is no
longer of its conformal form, the assumptions that undergird the
conformal computation break down: either conformal
invariance fails or the pointlike couping $G^2 \O$ develops
a form factor, in both cases acting to reduce the cross-section. 
As emphasized in \cite{ds09}, the
dominant cross-section for $gg\to 4\gamma$ is at very large $\hat s$,
because $d\sigma/d\hat s$ initially grows with $\hat s$ even after the
falling parton distribution functions are accounted for, shrinking
only at multi-TeV energies.  Thus for the cross-section to be
correctly computed, the energy at which conformal invariance
breaks down must be somewhat larger than the energy $\sm$ 
at which the cross-section peaks.  This constraint was
computed for a 10 TeV LHC in \cite{ds09}.  Here we 
use the self-consistency constraints for a 14 TeV LHC.

For smaller $\D_\g$, we need bounds on both scales.  We obtain
constraints on $\L_\g$ using our direct LEP II bounds on this quantity
at small $\D_\g$ from table \ref{table1}, and using unitarity considerations at
large $\D_\g$.  Since there are four photons in the final state, we
require consistency for all diphoton invariant masses up to $\sm/2$,
noting (see below) that the dominant cross-section arises where both photon
pairs have invariant mass of this order.

For $C_3$, constraints can be read off from Figures 1
and 2 of \cite{cr09}.  The absence of constraints for $\D_\g>1.7$ may
be purely technical, and perhaps other bounds may be obtained in this
region.  However, we only use the results of \cite{cr09} as they
currently stand.  

\begin{figure}[t]\centering
\includegraphics[width=0.25\textwidth]{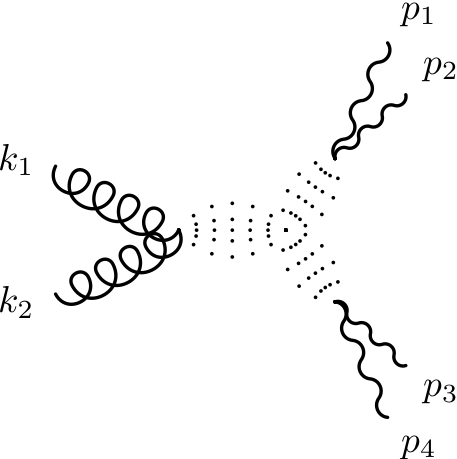}\hspace{1.5cm}
\includegraphics[width=0.25\textwidth]{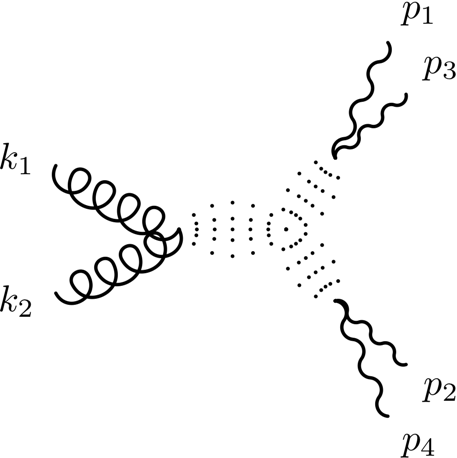}\hspace{1.5cm}
\includegraphics[width=0.25\textwidth]{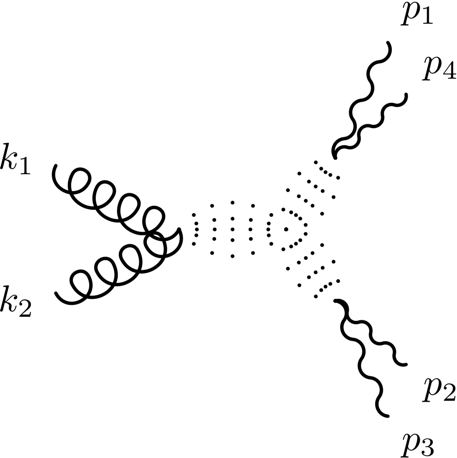}
\caption{Feynman diagrams for four photon production at the LHC. The dots represent the conformal three-point function.\label{feyn4photon}}
\end{figure}

We now combine these (for $\D_\g<1.7$)
with an overall bound on the squared matrix element, integrated
over phase space, allowing us to
obtain the results in table \ref{4photonbounds}.
In principle we could compute the exact cross-section (for a given $C_3$,
$\L_g$ and $\L_\g$,) but it is already
sufficient, as we will see, to make a rough estimate that bounds
the true cross-section from above.\footnote{More details will
be presented elsewhere.}
Since
\begin{equation}
\int d[{\rm Phase \ Space}] 
\ |{\cal M}|^2 < |{\cal M}^2|_{max} \int d[{\rm Phase \ Space}] \ ,
\end{equation}
and the phase space for four identical massless particles of total
energy $\sqrt{\hat s}$ can
be computed
\begin{equation}
\int d[{\rm Phase \ Space}] = \frac{1}{4!}\frac{\hat s^2}{2^{13} 3\pi^5}
\end{equation}
we only need to bound the squared matrix element.  We do this by bounding
${\cal M}$ itself, which contains three diagrams related by permutation of the
final state photons, as shown in figure \ref{feyn4photon}.  Let us consider the first diagram, where photons couple
to the hidden sector in pairs 1,2 and 3,4.  (The other diagrams give the same bound.)
The diagram factors into a standard model piece and a hidden sector piece.
The standard model piece can be bounded directly.   The kinematic
factor from the two gluons can be treated exactly, but
for the photons, with momenta
$p_1,p_2,p_3,p_4$, we make an approximation.
The two photon pairs each have a kinematic factor from
$F^{\mu\nu}F_{\mu\nu}$ which satisfies
\begin{equation}\label{photonkinematicfactor}
|\epsilon_i\cdot \epsilon_j\: p_i\cdot p_j - \epsilon_i\cdot p_j \:p_i\cdot \epsilon_j | <
p_i\cdot p_j = m_{ij}/2\ ,
\end{equation}
where $m_{ij}$ is the invariant mass of photons $i,j$.  
% begin figure
%end figure
Then we note that $m_{12}m_{34}$ times the hidden sector matrix element
can also be bounded; it is maximized where $m_{12}=m_{34}=\sqrt{\hat
  s}/2$.
Armed
with this bound on each of the three terms in the amplitude, we find
the partonic cross section at any $\sqrt{\hat s}$ is then bounded
by\footnote{Since there are three graphs in the amplitude, each of
  which has the same bound, there is an overall factor of $3^2$ in
  this expression, canceling the factors of 3 in the phase space
  integral. The existence of three diagrams appears to have been
  neglected in \cite{frt08}.  Inclusion would have increased rates,
  for a given $C_d$, by a factor of several, but would not much have
  affected the results quoted in \cite{frt08}, since the change
  affects both the Tevatron, where experimental bounds were obtained,
  and LHC, to which these bounds were extrapolated.}
\begin{equation}\label{sigmaApprox}
\hat\sigma<\frac{1}{2^{27}\pi^9}\frac{C_3^2}{\L_g^{2\D_g}\L_\g^{4\D_\g}}\hat s^{\D_g+2\D_\g-1}\Big[Q(\D_g,\D_\g)\Big]^2
\end{equation} 
with
\begin{equation}
Q(\D_g,\D_\g)=\frac{\G(4-\frac{\D_g}{2}-\D_\g)}{\G(2+\frac{\D_g}{2}-\D_\g)\big[\G(2-\frac{\D_g}{2})\big]^2}\int^1_0\!\!\!dx\!\!\int^{1-x}_0\!\!\!\!\!\!\!dy\frac{(xy)^{1-\D_g/2}(1-x-y)^{1+\D_g/2-\D_\g}}{\Big[xy+\frac{1}{4}(1-x-y)(x+y)\Big]^{4-\D_g/2-\D_\g}}.
\end{equation}

Finally, we integrate (\ref{sigmaApprox}) against the gluon-gluon
parton luminosity.\footnote{For technical reasons (calculational
speed) we have used the outdated CTEQ5M parton distribution functions
\cite{cteq5}.  As $gg$ luminosities are uncertain at high energies,
use of more up-to-date pdfs would shift our answers by up to a few
tens of percent.  This is comparable to other sources of uncertainty,
in particular the extraction of the minimum $\Lambda_g$ allowed by
Tevatron data and unitarity considerations. }
  At that point we need only substitute the appropriate
constraints on $\Lambda_g$, $\Lambda_\g$ and $C_3$ to obtain the
bounds displayed in table \ref{4photonbounds}.  

\subsection{Commentary}

In the table,
numbers shown in regular font are those for which
only experimental data was used.  For these, there is little ambiguity
and relatively small uncertainty.\footnote{Bounds on the $4\gamma$
  cross section obtained with purely experimentally-based constraints
  on the $\L_i$ are given in Appendix \ref{unitarityappendix}, in table
  \ref{boundwithoutunitarity}.  These bounds remain below a few fb for
  $\D_g+2\D_\g$ less than $\sim 4.4$.}  Numbers shown in boldface
are those for which unitarity considerations apply for either or both
$\L_g$ or $\L_\g$.  Theoretical uncertainties are somewhat larger here,
as much as a factor of 2.  Similar uncertainties apply for the numbers
in italics.  The relevant uncertainties in these regions are discussed
in \cite{ds09}.  It should be noted that it is possible to exceed these
bounds as long as one gives up conformal invariance; in this case the
rate could be larger, but is not predictable either in magnitude
or in its differential distributions.

It appears that the phenomenon suggested in \cite{frt08} is
unobservable at the LHC for smaller values of $\D_g$, $\D_\g$.  For
$\D_\g<1.7$ the rates are never better than marginal, and other
signals of a conformal hidden sector (such as jet-plus-MET or
two-photons-plus-MET) may be so much larger that they are easier to
observe despite larger backgrounds.  The weaker bounds for $\D_\g>1.7$
still allow for observable cross-sections, but it is quite possible
that there will eventually be bounds on $C_3$ in this regime.  (In the
special case studied in \cite{frt08} where the operators $\O_g$ and
$\O_\g$ to which the gluons and photons couple are the same operator,
the unitarity constraints of \cite{ds09} are more powerful, and the
numbers on the diagonal at $\D_g=\D_\g=1.8,1.9,2.0$ should be divided
by a factor \cite{ds09} of 33.) We emphasize also that most conformal
field theories do not saturate unitarity bounds.  We conclude that
four-photon production from unparticle interactions is unlikely to be
a discovery channel for a conformal hidden sector, or even an
observable signal in many cases.

Our work indicates that this direction of research
uncovers nothing surprising about
conformal field theory.  Naively, one would have expected that in a
hidden sector with no mass gap, the dominant signals would be in
channels with missing momentum, and that the cost to obtain a visible
signal would be high, leading only to relatively small and subtle
signals.\footnote{Similar
  naive intuition suggests that two-photon-plus-MET signals are almost
  always larger than the four-photon signals, because the latter is
  suppressed by $\Lambda_\gamma^{4\D_\g}$ while the former is
  suppressed only by $\Lambda_\gamma^{2\D_\g}$.  It is possible to
  prove that the four-photon signal can only exceed the
  digamma-plus-MET signal by a logarithmic enhancement, and this only
  in extreme circumstances.  We therefore suspect that any discovery
  of a hidden sector coupling to gauge bosons will occur in a MET
  signal, either with an ISR jet or with two photons.}
  (This is in contrast to ``hidden valleys'' \cite{hiddenvalley}
where, because of a mass gap in the hidden sector, the visible
signatures may easily and naturally dominate.)  The suggestion of
\cite{frt08} flies in the face of this expectation.  But in fact, the
naive intuition appears to be essentially correct.

\section{Conclusion}

We have considered bounds on couplings of scalar operators built from
electroweak bosons to hidden sectors with an exact or approximate
conformal invariance above a few GeV.  Such ``unparticle'' sectors are
significantly constrained by LEP I and LEP II data on
photon-plus-nothing events. We have provided constraints on couplings
to both $SU(2)$ and $U(1)$ gauge bosons for $1\leq \D_\O\leq 2$.  These
are particularly powerful at smaller values of $\D_\O$.  

We have also used these results, and those of
\cite{ds09} and \cite{cr09}, to constrain four-photon production at
the 14 TeV LHC, dramatically improving the bounds for $\D_\g$ in the
range 1 to 1.7 from of order several pb to far less than 5 fb.  For
$\D_\g$ near 2, where the bounds of \cite{cr09} are not available, the
best bounds (a femtobarn if $\O_g=\O_\g$, as in \cite{frt08}, and a
few tens of femtobarns in the more general case) still come from the
methods of \cite{ds09}, due to the lack of a bound on the three-point
OPE coefficient from \cite{cr09}.
It seems likely that these bounds
will be further strengthened as more is learned about the unitarity
constraints on conformal field theory.  In particular, the powerful
methods of \cite{cr09} may not yet have been exhausted,
and may yet give additional constraints at $\Delta_\g>1.7$.

It is also worth noting that constraints on $\L_g$ will sharply improve with
early data at the LHC.  By the time 1 inverse fb of data 
is obtained at the 14 TeV LHC, it seems likely, if no jet-plus-MET signal
is observed, that bounds on $\L_g$
will improve by a factor of 5 or so relative to the bounds at
the Tevatron.  This in turn will even further tighten limits on
four photon events, long before there is any chance of seeing them.  Conversely,
if a four-photon signal is observable at the LHC, it seems likely 
that a jet-plus-MET signal will be detected first.  

\section*{Acknowledgements}

We are grateful to C. Mateuzzi, K. Cranmer, Y. Gershtein, and
S. Somalwar for useful discussions. The work of S. Knapen was
partially supported by the Belgian American Educational Foundation and
the Franqui Foundation.  The work of M.J.S. was supported by NSF grant PHY-0904069
and by DOE grant DE-FG02-96ER40959.

\clearpage
\appendix
\section{Figures}

These figures summarize our results for experimental bounds on the strength of CFT coupling to the
electroweak gauge bosons. The values $\L_1$ and $\L_2$ are defined in equation \ref{firstlambdadef}.
The plots represent the allowed regions for these variables given constraints from LEP I and LEP II only.
The effect of interference between the photon and $Z$ channel for these graphs is very small, and they are
drawn for $\delta = 0$, where $\delta$ is defined in equation \ref{deltadef}. The graphs include all contributions 
from equation \ref{alambda}.

% Begin a list of figures
\begin{figure}[h!]
\begin{center}
\includegraphics[width=80mm,height=0.32\textheight]{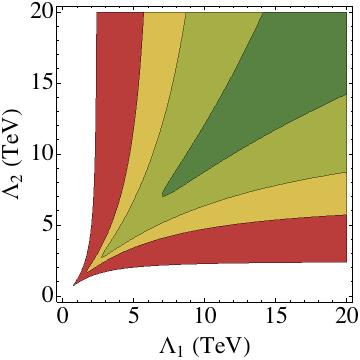}
\caption{Plot of 95\% CL allowed regions of $\L_1$ vs $\L_2$, in units of TeV, from LEP I data for 
$\delta = 0$. The shaded areas, from largest
area to smallest, are the allowed regions for $\D = 1.5, 1.35, 1.2, 1.05$. 
}
\label{fig1}
\end{center}
\end{figure}

\begin{figure}[h!]
\begin{center}
\includegraphics[width=80mm,height=0.32\textheight]{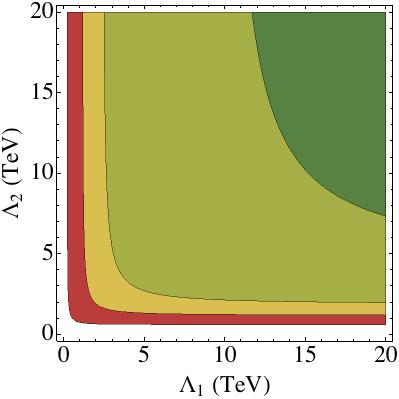}
\caption{Plot of 95\% CL allowed regions of $\L_1$ vs $\L_2$, in units of TeV, from LEP II data for 
$\delta =0$. The shaded areas, from largest
area to smallest, are the allowed regions for $\D = 1.5, 1.35, 1.2, 1.05$. 
}
\label{fig2}
\end{center}
\end{figure}

\begin{figure}[t]
\begin{center}
\includegraphics[width=80mm,height=0.32\textheight]{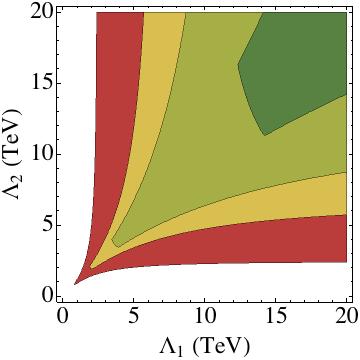}
\caption{Combined plot of 95\% CL allowed regions of $\L_1$ vs $\L_2$,
in units of TeV, from both LEP I and LEP II data. This represents
the combination of the two previous figures without careful
statistical weighting.
At the corners of the contours  (where both bounds saturate) the true
95\% contours would be more rounded than shown. The shaded areas,
from largest area to smallest, correspond to
$\D = 1.5, 1.35, 1.2, 1.05$. }
\label{fig3}
\end{center}
\end{figure}

\clearpage

\section{Bounds without unitarity argument\label{unitarityappendix}}

In the table below are shown the bounds on $gg\to\gamma\gamma\gamma\gamma$
that would be obtained with our
methods using only experimental bounds on $\L_g$ and $\L_\g$ and the 
unitarity bounds on the conformal three-point function coefficient
$C_3$ from \cite{cr09}.   No theoretical assumptions go into these
bounds, so they are particularly robust.

\begin{table}[h!]
\begin{tabular}{|c||r@{.}l|r@{.}l|r@{.}l|r@{.}l|r@{.}l|r@{.}l|r@{.}l|r@{.}l|}\hline
 $\phantom{a}_{\D_g} \diagdown^{\D_\g}$ & \multicolumn{2}{c|}{1.05} &\multicolumn{2}{c|}{ 1.1} &\multicolumn{2}{c|}{ 1.2} &\multicolumn{2}{c|}{ 1.3} &\multicolumn{2}{c|}{ 1.4} &\multicolumn{2}{c|}{ 1.5} & \multicolumn{2}{c|}{1.6} &\multicolumn{2}{c|}{ 1.7} \\\hline\hline
 1.05 & $2$&$7\times 10^{\text{-6}}$ & $2$&$7\times 10^{\text{-5}}$ & $4$&$8\times 10^{\text{-4}}$ & 0&010 & 0&093 & 0&62 & 6&4 & 110& \\\hline
 1.1  & $5$&$2\times 10^{\text{-6}}$ & $5$&$2\times 10^{\text{-5}}$ & $6$&$7\times 10^{\text{-4}}$ & 0&014 & 0&13 & 0&89 & 9&2 & 110 &\\\hline
 1.2  & $1$&$5\times 10^{\text{-5}}$ & $1$&$4\times 10^{\text{-4}}$ & $1$&$3\times 10^{\text{-3}}$ & 0&029 & 0&37 & 2&4 & 19& & 180& \\\hline
 1.3  & $3$&$7\times 10^{\text{-5}}$ & $3$&$1\times 10^{\text{-4}}$ & $4$&$8\times 10^{\text{-3}}$ & 0&058 & 0&76 & 4&9 & 40& & 370& \\\hline
 1.4  & $9$&$0\times 10^{\text{-5}}$ & $7$&$8\times 10^{\text{-4}}$ & $9$&$6\times 10^{\text{-3}}$ & 0&12 & 1&6 & 10& & 85& & 800& \\\hline
 1.5  & $2$&$5\times 10^{\text{-4}}$ & $2$&$0\times 10^{\text{-3}}$ & 0&030 & 0&35 & 3&2 & 21& & 210& & 1600& \\\hline
 1.6  & $6$&$1\times 10^{\text{-4}}$ & $5$&$2\times 10^{\text{-3}}$ & 0&060 & 0&71 & 6&6 & 44& & 520& & 4000& \\\hline
 1.7  & $1$&$9\times 10^{\text{-3}}$ & 0&014 & 0&17 & 1&9 & 18& & 110& & 1100& & 9600& \\\hline
 1.8  & $4$&$1\times 10^{\text{-3}}$ & 0&023 & 0&24 & 3&9 & 46& & 230& & 2600& & 23000& \\\hline
 1.9  & $7$&$3\times 10^{\text{-3}}$ & 0&040 & 0&49 & 6&0 & 77& & 600& & 6500& & 63000& \\\hline
\end{tabular}
\caption{Bounds, in fb, on 4 photon production at the LHC (14 TeV),
using only constraints from experiment and internal consistency of the
conformal field theory. No unitarity arguments are used here in constraining
$\L_g$ or $\L_\g$. There is
no bound for $\D_\g>1.7$, since no bound on $C_3$ is known
there. \label{boundwithoutunitarity}}
\end{table}

\section{Detail of analysis \label{analysisappendix}}

At both LEP I and LEP II, the highest signal-to-background ratio is found for large photon energies. The only significant source of standard model background in this regime is $\nu\bar\nu\gamma$ production, which falls quickly with rising photon energy \cite{background}. As mentioned before, the unparticle signal is peaked at higher photon energy. 

Optimal values for the photon energy cut were found using the
following method. For any potential value of the cut energy
$E_\textrm{cut}$, the bound on $\L_Z$ or $\L_\g$ was calculated for any
number $n$ of observed photon events with energy above the cut. These
bounds were then averaged over $n$, using a Poisson distribution for
$n$ assuming the only source is the background. The value of
$E_\textrm{cut}$ that produces the strongest average expected bound on
$\L_Z$ or $\L_\g$ via this method is then used as the energy cut,
following \cite{poisson}.

For LEP I, we can reproduce reasonably well the shape of each experiment's Monte
Carlo simulation of background with the background model
\cite{background}.  We fit the normalization 
to account for each experiment's efficiency.  An additional
complication was that L3 had a larger angular acceptance than OPAL,
DELPHI and ALEPH. For the purpose of being conservative, we only
considered the signal in the wedge that all four detectors have in
common, but took into account the background for the entire L3
detector.

The cut energies for different values of $\D$ were calculated to the nearest 0.2 GeV, maximizing the expected bound on $\L_Z$, and can be found in table \ref{lep1cuts}. The signal efficiencies for these cuts range from 0.98 for $\D$ close to 1 to 0.74 for $\D$ close to 2. 

In the case of LEP II, the background model is less clear-cut. The
background in L3 and ALEPH is very small, and the resolution of the
plots is insufficient to make a reliable estimate. We therefore chose
to omit any background from L3 and OPAL in our analysis.  For OPAL and
especially DELPHI, the background is more significant. However,
because of the rather low resolution of DELPHI`s inner detector wedge
(HPC), we cannot reproduce the shape of DELPHI`s Monte Carlo with the
background model of \cite{background}. Instead we used a more general
fit function with three fit parameters to model the background in OPAL
and DELPHI.

Furthermore, DELPHI is the only experiment that has separate plots available for the different segments of its detector. Since the signal is rather flat in $\cos\theta$ and the background is peaked in the forward region, we only consider DELPHI`s inner wedge to increase the signal to background ratio. As DELPHI`s resolution is inferior compared to the other three experiments, it has some background events leaking into the signal region, resulting from the smearing of the Z-peak. These events significantly weaken our bound for values of $\D$ close to 2.

To calculate the cuts, the same analysis was performed as was done for LEP I, but now in terms of missing mass. The cut on the missing mass was calculated to the nearest 1.0 GeV, maximizing the expected bound on $\L_\g$, and can be found in table \ref{lep1cuts}. 

\begin{table}[h]
\begin{tabular} {|c|c|c|c|c|c|c|c|c|c|c|c|c|}  \hline
$\D$&1&1.05&1.1&1.2&1.3&1.4&1.5&1.6&1.7&1.8&1.9&2\\\hline
$E_\textrm{cut}$ (GeV)&43.8&42.0&40.8&39.0&37.8&36.6&35.8&34.8&34.0&33.2&32.4&31.6\\\hline
\end{tabular}
\caption{The photon energy cut for the different values of $\D$ for LEP I.\label{lep1cuts}}
\end{table}

\begin{table}[h]
\begin{tabular} {|c|c|c|c|c|c|c|c|c|c|c|c|c|}  \hline
$\D$&1&1.05&1.1&1.2&1.3&1.4&1.5&1.6&1.7&1.8&1.9&2\\\hline
$M_\textrm{cut}$ (GeV)&15&22&26&30&33&35&37&39&41&43&45&48\\\hline
\end{tabular}
\caption{The missing mass cut for the different values of $\D$ for LEP II.\label{lep2cuts}}
\end{table}

To determine the bounds at 95$\%$ confidence level, the following equation was used, from \cite{poisson}
\begin{equation}\label{boundstat}
(1-0.95)\sum_{n=0}^{n_0}\frac{\mu_B^n}{n!} = e^{-N}\sum_{n=0}^{n_0}\frac{(\mu_B+N)^n}{n!},
\end{equation}
where $\mu_B$ is the expected number of background events, $n_0$ is the number of observed events, and $N$ is the 95$\%$ CL upper limit on the expected number of signal events. Note that if $n_0=0$, corresponding to no observed events, then $N=2.99$ independent of the number of expected background events.

At LEP I, none of the experiments observed any events with energies above the cuts. The bound was imposed by integrating the cross section \ref{photoncrosssection} above the appropriate energy cut, within the angular wedge that all four detectors shared ($\cos\theta< 0.7$), and accounting for the various detector efficencies and luminosities. The calculation was performed at each value of $\sqrt{s}$ used by the experiments,\footnote{We are thankful to K. Cranmer from L3 for providing us with this additional information.} and summed over all values, accounting for the various efficencies and luminosities.

At LEP II, some events were observed that passed the missing mass
cuts. The events were counted by hand from the graphs in
\cite{lep2opal,lep2delphi,lep2l3,lep2aleph}. As in LEP I, the
integrated cross section was bounded to the appropriate value computed
from equation \ref{boundstat}. Again, the calculation was performed at
each value of $\sqrt{s}$ used by the experiments, and summed over all
values, accounting for the various efficencies and luminosities. When
computing the expected signal, we have been conservative by only
accounting for angular acceptance that all detectors have in common
with DELPHI`s inner wedge ($\theta< 45^\circ$).

\providecommand{\href}[2]{#2}\begingroup\raggedright\endgroup


\begin{thebibliography}{10}

\bibitem{ds09}
A.~Delgado and M.~J. Strassler, \emph{{A Simple-Minded Unitarity Constraint and
  an Application to Unparticles}}, Phys. Rev. {\bf D81} (2010)
  \href{http://dx.doi.org/10.1103/PhysRevD.81.056003}{056003},
\href{http://arxiv.org/abs/0912.2348}{{\tt arXiv:0912.2348 [hep-ph]}}
%%CITATION = 0912.2348;%%.

\bibitem{cr09}
F.~Caracciolo and S.~Rychkov, \emph{{Rigorous Limits on the Interaction
  Strength in Quantum Field Theory}},
\href{http://arxiv.org/abs/0912.2726}{{\tt arXiv:0912.2726 [hep-th]}}
%%CITATION = 0912.2726;%%.

\bibitem{frt08}
J.~L. Feng, A.~Rajaraman  and H.~Tu, \emph{{Unparticle self-interactions and
  their collider implications}}, Phys. Rev. {\bf D77} (2008)
  \href{http://dx.doi.org/10.1103/PhysRevD.77.075007}{075007},
\href{http://arxiv.org/abs/0801.1534}{{\tt arXiv:0801.1534 [hep-ph]}}
%%CITATION = 0801.1534;%%.

\bibitem{Georgi:2010zz}
H.~Georgi, \emph{{Unparticle physics}}, Int. J. Mod. Phys. {\bf A25} (2010)
\href{http://dx.doi.org/10.1142/S0217751X1004886X}{573--586}
%%CITATION = IMPAE,A25,573;%%.

\bibitem{gir08}
B.~Grinstein, K.~A. Intriligator  and I.~Z. Rothstein, \emph{{Comments on
  Unparticles}}, Phys. Lett. {\bf B662} (2008)
  \href{http://dx.doi.org/10.1016/j.physletb.2008.03.020}{367--374},
\href{http://arxiv.org/abs/0801.1140}{{\tt arXiv:0801.1140 [hep-ph]}}
%%CITATION = 0801.1140;%%.

\bibitem{nelson}
A.~E. Nelson, M.~Piai  and C.~Spitzer, \emph{{Protecting unparticles from the
  MSSM Higgs sector}}, Phys. Rev. {\bf D80} (2009)
  \href{http://dx.doi.org/10.1103/PhysRevD.80.095006}{095006},
\href{http://arxiv.org/abs/0905.0503}{{\tt arXiv:0905.0503 [hep-ph]}}
%%CITATION = 0905.0503;%%.

\bibitem{cky07}
K.~Cheung, W.-Y. Keung  and T.-C. Yuan, \emph{{Collider Phenomenology of
  Unparticle Physics}}, Phys. Rev. {\bf D76} (2007)
  \href{http://dx.doi.org/10.1103/PhysRevD.76.055003}{055003},
\href{http://arxiv.org/abs/0706.3155}{{\tt arXiv:0706.3155 [hep-ph]}}
%%CITATION = 0706.3155;%%.

\bibitem{fds03}
B.~Field, S.~Dawson  and J.~Smith, \emph{{Scalar and pseudoscalar Higgs boson
  plus one jet production at the LHC and Tevatron}}, Phys. Rev. {\bf D69}
  (2004) \href{http://dx.doi.org/10.1103/PhysRevD.69.074013}{074013},
\href{http://arxiv.org/abs/hep-ph/0311199}{{\tt arXiv:hep-ph/0311199}}
%%CITATION = HEP-PH/0311199;%%.

\bibitem{lep1delphi}
{\bf {DELPHI}} Collaboration, P.~Abreu {\em et al.}, \emph{{Search for new
  phenomena using single photon events in the DELPHI detector at LEP}}, Z.
  Phys. {\bf C74} (1997)
\href{http://dx.doi.org/10.1007/s002880050421}{577--586}
%%CITATION = ZEPYA,C74,577;%%.

\bibitem{lep1l3}
{\bf {L3}} Collaboration, M.~Acciarri {\em et al.}, \emph{{Search for new
  physics in energetic single photon production in $e^{+} e^{-}$ annihilation
  at the $Z$ resonance}}, Phys. Lett. {\bf B412} (1997)
\href{http://dx.doi.org/10.1016/S0370-2693(97)01003-4}{201--209}
%%CITATION = PHLTA,B412,201;%%.

\bibitem{lep1opal}
{\bf {OPAL}} Collaboration, R.~Akers {\em et al.}, \emph{{Measurement of single
  photon production in e+ e- collisions near the Z0 resonance}}, Z. Phys. {\bf
  C65} (1995)
\href{http://dx.doi.org/10.1007/BF01571303}{47--66}
%%CITATION = ZEPYA,C65,47;%%.

\bibitem{lep1aleph}
{\bf {ALEPH}} Collaboration, D.~Buskulic {\em et al.}, \emph{{A Direct
  measurement of the invisible width of the Z from single photon counting}},
  Phys. Lett. {\bf B313} (1993)
\href{http://dx.doi.org/10.1016/0370-2693(93)90027-F}{520--534}
%%CITATION = PHLTA,B313,520;%%.

\bibitem{poisson}
S.~I. {Bityukov} and N.~V. {Krasnikov}, \emph{{Confidence intervals for the
  parameter of Poisson distribution in presence of background}}, ArXiv Physics
  e-prints (Sept., 2000)
\href{http://arxiv.org/abs/arXiv:physics/0009064}{{\tt arXiv:physics/0009064}}
% .

\bibitem{lep2delphi}
{\bf {DELPHI}} Collaboration, J.~Abdallah {\em et al.}, \emph{{Photon events
  with missing energy in e+ e- collisions at s**(1/2) = 130-GeV to 209-GeV}},
  Eur. Phys. J. {\bf C38} (2005)
  \href{http://dx.doi.org/10.1140/epjc/s2004-02051-8}{395--411},
\href{http://arxiv.org/abs/hep-ex/0406019}{{\tt arXiv:hep-ex/0406019}}
%%CITATION = HEP-EX/0406019;%%.

\bibitem{lep2aleph}
{\bf {ALEPH}} Collaboration, A.~Heister {\em et al.}, \emph{{Single photon and
  multiphoton production in $e^{+} e^{-}$ collisions at $\sqrt{s}$ up to
  209-GeV}}, Eur. Phys. J. {\bf C28} (2003)
\href{http://dx.doi.org/10.1140/epjc/s2002-01129-7}{1--13}
%%CITATION = EPHJA,C28,1;%%.

\bibitem{lep2l3}
{\bf {L3}} Collaboration, P.~Achard {\em et al.}, \emph{{Single photon and
  multiphoton events with missing energy in $e^{+} e^{-}$ collisions at LEP}},
  Phys. Lett. {\bf B587} (2004)
  \href{http://dx.doi.org/10.1016/j.physletb.2004.01.010}{16--32},
\href{http://arxiv.org/abs/hep-ex/0402002}{{\tt arXiv:hep-ex/0402002}}
%%CITATION = HEP-EX/0402002;%%.

\bibitem{lep2opal}
{\bf {OPAL}} Collaboration, G.~Abbiendi {\em et al.}, \emph{{Photonic events
  with missing energy in e+ e- collisions at s**(1/2) = 189-GeV}}, Eur. Phys.
  J. {\bf C18} (2000) \href{http://dx.doi.org/10.1007/s100520000522}{253--272},
\href{http://arxiv.org/abs/hep-ex/0005002}{{\tt arXiv:hep-ex/0005002}}
%%CITATION = HEP-EX/0005002;%%.

\bibitem{cdf}
{\bf {CDF}} Collaboration, T.~Aaltonen {\em et al.}, \emph{{Search for large
  extra dimensions in final states containing one photon or jet and large
  missing transverse energy produced in $p \bar{p}$ collisions at $\sqrt{s}$ =
  1.96-TeV}}, Phys. Rev. Lett. {\bf 101} (2008)
  \href{http://dx.doi.org/10.1103/PhysRevLett.101.181602}{181602},
\href{http://arxiv.org/abs/0807.3132}{{\tt arXiv:0807.3132 [hep-ex]}}
%%CITATION = 0807.3132;%%.

\bibitem{cteq5}
{\bf {CTEQ}} Collaboration, H.~L. Lai {\em et al.}, \emph{{Global QCD analysis
  of parton structure of the nucleon: CTEQ5 parton distributions}}, Eur. Phys.
  J. {\bf C12} (2000) \href{http://dx.doi.org/10.1007/s100529900196}{375--392},
\href{http://arxiv.org/abs/hep-ph/9903282}{{\tt arXiv:hep-ph/9903282}}
%%CITATION = HEP-PH/9903282;%%.

\bibitem{hiddenvalley}
M.~J. Strassler and K.~M. Zurek, \emph{{Echoes of a hidden valley at hadron
  colliders}}, Phys. Lett. {\bf B651} (2007)
  \href{http://dx.doi.org/10.1016/j.physletb.2007.06.055}{374--379},
\href{http://arxiv.org/abs/hep-ph/0604261}{{\tt arXiv:hep-ph/0604261}}
%%CITATION = HEP-PH/0604261;%%.

\bibitem{background}
G.~Barbiellini {\em et al.}, \emph{{NEUTRINO COUNTING}},
Presented at Workshop on Z Physics at LEP
% .

\end{thebibliography}
\end{document}